# Design Framework for Conversational Agent in Couple relationships: A Systematic Review

Soyoung Jung[1]  and Sung Park[2]

**Abstract**
The development of conversational agents (CA) has demonstrated promising potential in supporting users' mental health through dialogue. While numerous studies have explored CA design for individual psychological support, systematic investigations into CAs designed to assist couples experiencing relational and psychological difficulties remain limited. The present study aims to identify design considerations for conversational agent interventions that reflect the relational context of couples and support their mental and emotional well-being. Following the PRISMA guidelines, a systematic review was conducted across seven major databases -- CINAHL, Embase, PubMed, PsycINFO, Scopus, Web of Science, and the ACM Digital Library. Only peer-reviewed empirical studies were included, with duplicates removed and selection criteria applied, resulting in a final set of twelve studies for analysis. A thematic analysis was conducted across three analytical dimensions: AI interaction design, relational framing, and technical limitations. Three overarching themes emerged from the synthesis: (1) the need to design a relational expert persona, (2) technological design directions that leverage state-of-the-art (SOTA) technologies to support relational specificity and emotional competence in couples, and (3) a paradigm shift from content-centered to relationship-centered design. Based on these themes, the study proposes eight key design considerations for couple-oriented CAs: (1) agent persona, (2) individual mode, (3) concurrent mode, (4) conjoint mode, (5) ethics, (6) data and privacy, (7) interaction pattern, and (8) safety mechanism. These considerations outline directions for designing CAs as relationally sensitive mediators that maintain multiple alliances, respect cultural and ethical boundaries, and ensure fairness and emotional safety between partners. Ultimately, this systematic review proposes a design framework that integrates relational theory with cutting-edge AI technologies to guide the development of conversational agents for couple-based mental health interventions.

## 1. Introduction

Couples in close relationships are not only affected relationally when conflicts arise but also experience direct consequences on individual well-being, reflecting their inherently systemic interdependence. Relationship difficulties extend beyond separation or divorce; many couples remain together while harboring profound dissatisfaction. Notably, approximately 75% of individuals entering personal therapy report relationship stress as a primary concern, underscoring the deep interconnection between individual mental health and relational context (Sperry, 2018). Thus, couple distress operates within an interdependent structure in which one partner's decline in mental health adversely affects the other's emotional well-being and overall relationship satisfaction (Landolt et al., 2023). Accordingly, interventions targeting couple relationships need to support individual well-being through diverse and adaptive approaches. However, most couple-based interventions remain confined to traditional in-person or telephone formats, limiting their scalability due to constraints related to accessibility, cost, and time (Gil et al., 2024).

[1]  1st author: Department of Family Welfare, Sangmyung University, Seoul, Republic of Korea;
- Email address: 202231018@sangmyung.kr   - ORCID number: 0009-0005-9498-5123

[2]  Corresponding author: School of Data Science and Artificial Intelligence, Taejae University, Seoul, Republic of Korea
- Email: sjp@taejae.ac.kr - ORCID number: 0000-0002-1242-9264

To address these limitations, advances in AI technologies have introduced new possibilities. Conversational agents (CA) -- systems capable of engaging in human-like verbal interaction -- have enhanced the accessibility and sustainability of mental health care (Ahmad et al., 2022). Recent meta-analyses further demonstrate that CA interventions significantly reduce depression and stress, highlighting their potential not only as accessible tools but also as effective therapeutic modalities (Li et al., 2023). However, most existing CA research remains focused on individual-centered support, which inadequately captures the emotional dynamics and relational processes that emerge within couple interactions. Although a few studies have applied CAs to couple contexts, systematic evidence on how to design such agents in ways that reflect relational dynamics remains limited.

This study aims to synthesize existing literature to identify key design considerations for conversational agent interventions targeting couples and to propose design principles that can support mental health enhancement and relationship maintenance strategies.

**2. Rationale**

When designing interventions for couples within relational contexts, it is essential to look beyond technical or methodological concerns and examine the underlying criteria that should guide such efforts. To address this question, the present study integrates two complementary perspectives: first, identifying the distinctive characteristics of couple relationships that constitute the target of intervention, and second, determining the design principles by which conversational agents should be developed to effectively engage within these relationships.

*Couple: What to Consider*

Couples are not merely two individuals coexisting but a relational system in which emotions and behaviors are reciprocally intertwined (Lebow & Snyder, 2022). Accordingly, the design of couple interventions should not focus on resolving the problems of one individual but rather on addressing the shared emotional context co-constructed by both partners. In this regard, the common factors proposed by Sprenkle, Davis, and Lebow (2009) offers critical insight: therapeutic change is determined less by specific techniques than by the quality of relational interactions, namely, the responsiveness and feedback occurring within the system. The foremost consideration in designing interventions for couples lies in adopting a theoretical sensitivity that reflects the relational context. The presence or absence of a relational framework not only shapes interpretation but also determines the overall direction of the intervention. For instance, whereas a problem-focused approach views conflict as an issue to be solved, a systemic perspective interprets it as a signal of interactional patterns and prompts exploratory questions about its underlying context. This serves as a foundational basis for designing relational questions.

Furthermore, the scope of couple-focused interventions extends beyond relationship maintenance. In practice, such interventions encompass a broad range of themes, including relationship enhancement, post-relationship adjustment following divorce or bereavement, intimate partner violence (IPV) prevention, reconciliation, separation, and decision-making processes (Sperry & Peluso, 2018). Therefore, interventions addressing couple relationships should not aim solely at restoring happiness but rather encompass the full life-span spectrum of relational experiences. Designers must recognize the specific stage of the couple's relationship targeted by the intervention and construct dialogue structures that align with its corresponding needs and dynamics.

Finally, relational interventions are inherently characterized by multilayered treatment modes. These can be classified according to whether the intervention targets individuals reflecting on their relationship (individual mode), partners receiving treatment separately but from the same therapist (concurrent mode), or partners engaging in therapy together with the same therapist (conjoint mode). Understanding these treatment modes is essential for determining how a conversational agent should adapt its interaction strategies. In this context, the core of couple interventions lies in distinguishing the appropriate level of dialogue required for each situation, illustrating the potential for conversational systems to move beyond single-user interactions.

*Conversational Agent: What to Consider*
CAs are systems designed to interact with humans through text, speech, and nonverbal signals (McTear, 2016). However, designing a CA for relational contexts such as couples extends beyond the technical challenge of achieving natural dialogue. It involves addressing how artificial intelligence can understand human emotions and mediate interpersonal relationships.

This study focuses on human-like competence as the core of CA design. This competence comprises cognitive, emotional, and relational capacities, among which relational competence plays a particularly vital role in fostering user engagement within psychological and mental health contexts (Chandra, Shirish, & Srivastava, 2022). Relational competence refers to the ability to recognize and interpret the emotions and intentions of others and to respond with attuned and appropriate attitudes.

Designing for human-like competence requires attention to how users perceive the agent, not merely as a tool, but as a social actor (Nißen et al., 2022). In domains such as mental health, where psychological support and trust are essential, defining the agent's role and persona becomes particularly important. The design of an AI persona involves establishing human-like attributes such as a name, personality, and role (Nielsen, 2014). Such explicit role definition helps prevent relational ambiguity and ensures coherence throughout the interaction. Therefore, this study examines how persona design contributes to realizing human-like competence in conversational agents.

Within human-like competence, emotional capacities such as emotion detection and appropriate expression are even more essential than cognitive abilities. Emotion recognition is not merely a process of labeling but a means of building relational trust through contextually appropriate empathic responses (Allouch et al., 2021). Furthermore, emotional competence is closely tied to contextual understanding. Thus, the inclusion of emotion recognition, emotional expression, and contextual awareness represents a core consideration in the design of conversational agents.

In the design of conversational agents intended for psychological interventions, transparency and safety must be prioritized. Given that dialogues within couple interventions naturally involve sensitive and personal information, it is essential to disclose how such data are processed and used, thereby establishing clear boundaries and safeguarding users' psychological autonomy (Allouch et al., 2021; Ahmad et al., 2022).

## 3. Research Questions
Although Conversational AI technologies have advanced considerably, most applications remain focused on individuals. There is a lack of systematic and comprehensive research that specifically addresses the couple relationship context. Therefore, this study aims to conduct a systematic review following PRISMA guidelines of empirical studies that implemented conversational AI Agents for couple relationship interventions.

The study seeks to summarize how conversational interactions between AI agents and users were designed, how relational contexts were considered and reflected, what technologies were employed to ensure smooth dialogue.

The research questions (RQs) are as follows:
RQ1: How do AI agents reflect the context of couples interacting with users?
RQ2: In what ways is the relational context incorporated?
RQ3: How are conversational technologies implemented to support interventions?
Final RQ: Synthesizing RQ1-RQ3, what are the key design considerations for conversational AI Agents within the couple context?

For a more detailed literature review, the study categorizes the research questions into four dimensions: 1) AI interaction design 2) Relational Framing 3) Implementation 4) User experience insights. In addition, the specific definitions of each dimension and their sub-dimension of review are presented in the Related works (See Table 1).

*[Table 1. Design Rationale of Dimension and Sub-dimensions by RQs]*

| Dimension | Sub Dimension | Design Rationale |
|---|---|---|
| RQ1: CA interaction design | Role-based persona | Role-based persona (Nielsen, L., 2014) |
| | Naming/Personality | Relational Competence: Human-like persona (Chandra, S. et al., 2022) |
| | Context-Aware | Good conversation (Clark, L., et al. 2019) |
| | Consideration of Transparency* (Safety and privacy issue) | Healthcare domain challenges (Allouach, et al., 2021) *(Ahmad, R., et al. 2022) |
| RQ2: Relational Framing | Application of Relational theoretical basis | Reflection of the relational context (Sprenkle, Davis, and Lebow, 2009) |
| | Definitions of Relational problems/ relationships Topics | relationship enhancement, post-relationship adjustment following divorce or bereavement, intimate partner violence (IPV) prevention, reconciliation, separation, and decision-making (Sperry, L., & Peluso, P., 2018) |
| | Treatment modes | Individual, Concurrent, Conjoint (Sperry, L., & Peluso, P., 2018) |
| RQ3: implementation | Emotional competence (detection/expression) | Human-centered conversation (Allouach et al., 2021) |
| | Responds to user's emotion expression | Facilitating user engagement in the social and health domain (Allouach et al., 2021) |
| | Interaction mode | The mode of communication (Allouach, et al., 2021) |
| | technologies to deliver dialogue | (Allouach, et al., 2021) |

## 4. Methods

The objective of this study is to comprehensively review existing literature to propose a design framework that reflects the couple relationship context. In particular, the study divides dimensions for detailed reviews according to the research questions and specifies the definitions and sub-dimensions through Related Works. The review process follows the PRISMA 2020 guidelines (Page et al., 2021).

### 4.1. Search Strategy

Based on fields of clinical and psychological sciences as well as AI design and engineering, the search was conducted in the following databases: ACM, CINAHL, Embase, PsyArticles, PsycINFO, PubMed, Scopus, and Web of Science. The search terms used involved combining three categories of terms through conjunctions. To maximize retrieval, each category included synonymous terms, and superscripts were used in developing the terms:
  (1) Couple terms: ("couple" OR "partner*" OR "spous*" OR "marital*" OR "romantic*" OR "intima*" OR "dyadic relationship")
  (2) Relationship contents: ("enhancement" OR "relationship enhancement" OR "communication" OR "mental health" OR "conflict" OR "emotion*" OR "resolution" OR "therapy" OR "counsel*" OR "support*")

    (3) Conversational agent technologies: ("conversational agent" OR "dialogue agent" OR "chatbot" OR "virtual agent" OR "embodied agent" OR "relational agent" OR "empathic agent" OR "AI agent" OR "embodied conversational agent" OR "voice assistant" OR "speech agent" OR "voice agent")

### 4.2. Eligibility Criteria

Studies were included if they met the following criteria:

① Research utilizing a conversational agent

② Research targeting couples or addressing couple relationships.

③ Empirical studies

④ Published in English

⑤ Peer-reviewed papers.

The exclusion criteria were as follows:

① Studies not peer-reviewed.

② Non-empirical studies

③ Studies not involving a conversational AI Agent

④ Studies without interaction between the conversational AI Agent and the user

⑤ Studies use conversational technologies but without addressing the couple relationship context in their interventions.

⑥ Non-English publications

⑦ Abstract-only papers

⑧ Master's theses or lower-level dissertations.

### 4.3. Data Collection Process

Data was organized using Excel matrices, while Mendeley Reference Manager was used for merging and extracting references. Screening was initially conducted by the first author based on titles and abstracts, and final inclusion decisions were made in consultation with another researcher to ensure reliability. In cases where the reflection of the couple relationship context was unclear, Marital and couple relationships expert consultation was sought.

### 4.4. Data analysis

This study aimed to examine the key characteristics of the selected studies according to each RQ and to synthesize them to explore the main design considerations within the context of couple relationships. The analysis followed an inductive (bottom-up) thematic analysis approach as outlined by Braun and Clarke (2006). While the procedure was grounded in their six-phase framework, the present analysis additionally reviewed the intersections among design-related dimensions across studies to identify relevant design factors.

    The analytic procedure consisted of six steps: First, during data extraction, study features, main findings, and UX-related statements were extracted from the included papers based on the RQs. Second, initial coding was conducted by coding the extracted content into core meaning units. Third, in searching for themes, the study characteristics and codes were compared and integrated to identify preliminary themes based on recurrent patterns. Fourth, during reviewing themes, the themes and codes were examined to ensure consistency and to identify differences across studies. Fifth, the themes were defined and named by specifying their meanings and labeling them in reflection of design considerations within the couple relationship context. Finally, writing up involved compiling the thematic findings and their design implications into the results sections.

To strengthen the thematic grounding, UX statements, major findings, and implications were jointly considered to interpret the features of conversational agents across research questions. However, variations were observed among the studies, some design-focused papers provided rich user statements, whereas others did not, and most tended to emphasize positive evaluations. Furthermore, while a few studies included direct user quotations, others relied on researcher-reported summaries, introducing potential bias. To ensure analytical balance and validity, both positive and negative evaluations, user requirements, and neutral comments were considered. These data were not treated as primary evidence but rather as contextual cues to support the interpretive process.

## 5. Results

A total of 1,295 records were retrieved from the eight databases using the search strategy. After removing 434 duplicates through Mendeley's merge function, 861 records remained, spanning studies from 1989 to 2025. Title and abstract screening were then conducted. According to the eligibility criteria, the following were excluded: 5 non-peer-reviewed studies, 43 non-empirical studies, 203 studies not involving conversational AI agents, studies without interaction between agent and user(e.g., focusing on only ethics, privacy, or self-disclosure without conversational intervention), and 400 intervention studies not considering couple relationships. As a results, 20 studies were included for full-text screening.

From the 20 studies subjected to full-text screening, 4 non-empirical studies and 2 studies that did not reflect the couple relationship context were excluded. Consequently, 12 studies were finally assessed (see Figure 1).

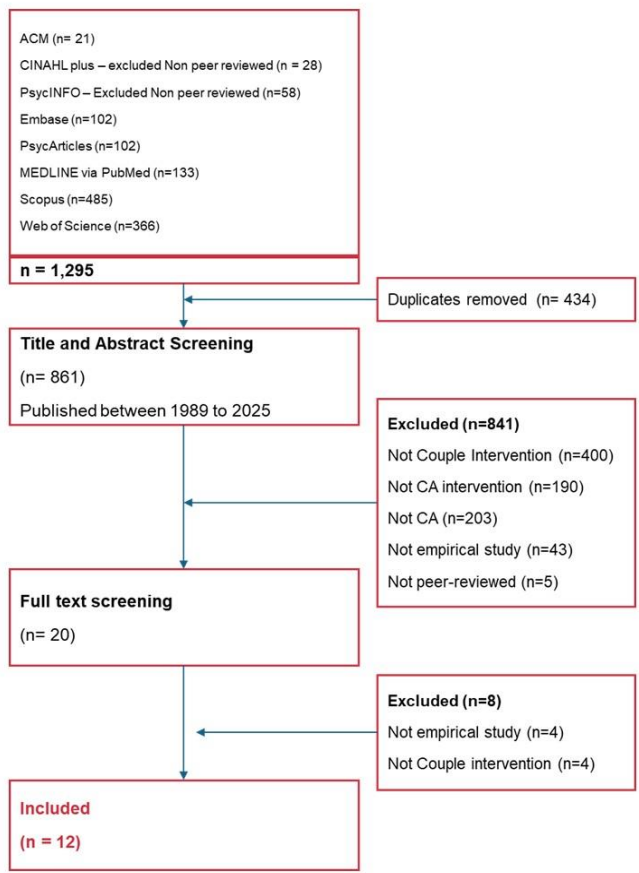

**Figure 1. Flow chart**

## 5.1. Study Characteristics

Using the 12 studies that were finally accessed, study characteristic was organized according to each RQs. Among the included studies, Schafer, Lachman, Zinser, et al. (2023) and Schafer, Lachman, Gardner, et al. (2025) employed the same CA design while conducting separate empirical studies. Although the study the by Van Velsen et al. (2024) satisfied the eligibility criteria, however, it provided limited information regarding the CA design. To supplement this description, we referred to its preceding study, Van Velsen et al. (2020), which had been excluded at the full-text screening stage as a non-empirical study. Li and Rau (2021) and De Filippo et al. (2023) modified their CA conditions for experimental purposes; hence, the condition most relevant to this review was selected for analysis. Specifically, the IoT-CA developed by Li and Rau (2021) represented a content-relevant use condition, whereas ChattyCuz introduced by De Filippo et al. (2023) reflected a gamified conversational agent. All included CAs were categorized based on the design characteristics implemented in each respective study.

### 5.1.1. Design dimension 1. AI Interaction Design

*Persona Setting*
In the context of psychological and emotional support, one of the key design elements for enhancing user engagement with AI agent is persona design. Among the persona elements, Role is particularly significant. We focused primarily on the name, personality, and role of the CA. Across the 12 studies, persona design was only partially reflected. We examined both the naming and personality traits of the agents, while roles were identified by coding statements in the original texts that implied role functions, even when not explicitly specified.

In terms of naming and personality settings, some studies assigned only a name without specifying any personality traits (Troitskaya & Batkhina, 2022; Van Velsen et al., 2024; Njogu et al., 2023). Conversely, other studies did not provide a name but reflected distinct personality characteristics, such as empathic (Yuksel & Kocaballi, 2022), unbiased (Zheng et al., 2021), or approachable, supportive, and friendly like an old cousin (De Filippo et al., 2023). Only one study incorporated both a name and a personality (Khatra et al., 2024). However, even in this case, the agent was not developed through a detailed persona design process. Likewise, several other studies did not attribute either a name or a personality to the agent (Li & Rau, 2021; Oshio, 2024; Mauricio et al., 2024; Schafer, Lachman, Gardner, et al., 2023; Schafer, Lachman, Zinser, et al., 2025).

Role was classified into two levels: a) roles according to the application of the CA (Allouch et al., 2021) and b) Roles as personas, that is, the ways in which CAs performed as agents (Nielsen, 2014). In terms of applications roles, 11 of the 12 studies could be classified as social-support agents, with the exception of Li & Lau's (2021) IoT-CA, which was positioned more as a General/open domain CA designed to explore communication satisfaction in the couple context rather than providing explicit social support.

Regarding role as personas, some studies explicitly stated the role, while in most cases it was not clearly mentioned. In such instances, we extracted relevant statements from the original texts based on the agent's stated objectives to identify the primary role(See Table 2). Overall, all agents were designed to facilitate and support user interactions, though specific functions varied across three major roles:

First, Expert Role provided interventions based on theoretical knowledge or evaluative criteria. For instance, Troitskaya and Batkhina (2022) conceptualized their chatbot as a "digital psychologist." Yuksel and Kocaballi (2022) guided users through CBT processes grounded in Ellis's ABC model. Mauricio et al. (2024) designed AyudaMujer to assess users' risk levels and connect them with professional services. Oshio (2024) created a divorce negotiation chatbot that offered legal information while also providing emotional support.

Second, Mediator Role is directly mediating interaction between partners. For example, Zheng et al.'s (2021) PocketBot delivered and moderated partner messages, allowing couples who struggled with direct expression to communicate more effectively. Khatra et al. (2024) designed *Anna* to facilitate relational connections by sharing smartphone usage data and encouraging partner participation.

Third, Guidance Role is guiding users through content or processes step by step and providing educational support. Van Velsen et al.'s (2024) LEAVES agent guided users through a bereavement journey, supporting mood check and offering coping strategies when grief became overwhelming. Schafer, Lachman, Gardner et al. (2023) and Schafer, Lachman, Zinser et al. (2025) developed ParentText, which automatically delivered parenting and violence-prevention messages. Li & Rau's (2021) IoT-CA participated in couple conversations by asking questions and providing information to guide conversation. Njogu et al.'s (2023) Nena delivered content and navigation via numeric responses, while De Filippo et al.'s (2023) ChattyCuz guided young women through gamified educational content.

In summary, the application role of most agents was that of social-supporting agents, while their persona roles consistently emphasized facilitating and supporting user experiences. More specifically, the primary roles identified were the Professional Role, Mediator Role, and Guidance Role.

*Context-Aware Dialogue*
For effective dialogue between AI agents and users, it is essential that conversations are remembered and reflected upon. CA should be able to memorize prior interactions and incorporate them into subsequent dialogues so as not to create discomfort in the user experience. The review of context-aware dialogue revealed that a few studies reflected contextual information only within a single conversational cycle, without mentioning processes of memory or carry-over to next cycles (Troitskaya & Batkhina, 2022; Yuksel & Kocaballi, 2022; Li & Rau, 2021; Oshio, 2024). In contrast, many studies made no reference to contextual reflection at all, and instead focused primarily on one-off functions such as evaluation and information delivery rather than sustaining the flow of interaction through dialogue (Zheng et al., 2021; Van Velsen et al., 2024; Mauricio et al., 2024; Schafer, Lachman, Zinser, et al., 2025; De Filippo et al., 2023; Njogu et al., 2023; Schafer, Lachman, Gardner, et al., 2023; Khatra et al., 2024).

*Consideration of Transparency*
Consideration of transparency, particularly regarding security and privacy design, varied across studies. In most cases, no explicit mention was made (Troitskaya & Batkhina, 2022; Yuksel & Kocaballi, 2022; Li & Rau, 2021; Van Velsen et al., 2024; Oshio, 2024; Mauricio et al., 2024). However, a few studies incorporated explicit strategies to safeguard personal information. For instance, PocketBot employed command-based design (e.g., "/") to prevent unintended disclosure of conversations (Zheng et al., 2021), while Anna included mechanisms to block personal data leakage between partners (Khatra et al., 2024). ParentText incorporated privacy protections though it was part of its research design, such as selective intake procedures, non-storage of sensitive responses, and anonymization of data (Schafer, Lachman, Zinser, et al., 2025; Schafer, Lachman, Gardner, et al., 2023). Similarly, ChattyCuz was designed with privacy as a central principle (De Filippo et al., 2023), and Nena prevented users from entering names or personal details (Njogu et al., 2023).

### 5.1.2. Design Dimension 2. Relational Framing

*Main relational topics*
For interventions in couple relationships, it is crucial to examine how relational problems are defined. The included studies addressed three main relational topics.

   The first topic was relationship improvement. Some studies targeted the dynamics of negative interactions (Troitskaya & Batkhina, 2022; Yuksel & Kocaballi, 2022), while others focused on reducing communication gaps in long-distance relationships (Zheng et al., 2021). Li & Rau (2021) and Khatra et al. (2024) emphasized participation in dialogue and activation of interaction. Schafer et al. (2023; 2025) developed ParentText, which provided parenting education to prevent IPV and thereby indirectly supported couple relationship improvement. In addition, Njogu et al. (2023) designed a sexual health chatbot for young Kenyan adults that addressed not only individual needs but also partner-related sexual satisfaction.

   The second topic was support after relational dissolution. LEAVES addressed bereavement following partner loss (Van Velsen et al., 2024), while Oshio (2024) provided both legal and emotional support in the process of divorce negotiation.

   The third topic was prevention of intimate partner violence (IPV). This included support for women at risk of IPV (Mauricio et al., 2024), as well as fostering gender equality attitudes and IPV prevention in young women's relationships (De Filippo et al., 2023).

*Application of Relational Theories*
Couple relationship intervention should be considered to apply relational perspectives. Among the included studies, only Troitskaya and Batkhina (2022) explicitly applied psychotherapeutic models that reflect the couple's relational aspects such as CBCT, IBCT, and EFT. Although not relational theories per se, some studies adopted general psychotherapeutic approaches: for example, Yuksel and Kocaballi (2022) and Van Velsen et al. (2024) applied CBT in their interventions. The majority of the other studies did not explicitly apply any relational theory (Li & Rau, 2021; Khatra et al., 2024; Oshio, 2024; Mauricio et al., 2024; Schafer et al., 2023, 2025; De Filippo et al., 2023; Njogu et al., 2023).

*Treatment Modes*
When addressing the couple context, interventions can be delivered in three modes: individual, concurrent, conjoint. In most of the assessed studies, interventions were conducted in the individual mode, where only one partner participated (Troitskaya & Batkhina, 2022; Yuksel & Kocaballi, 2022; Van Velsen et al., 2024; Oshio, 2024; Mauricio et al., 2024; De Filippo et al., 2023; Njogu et al., 2023). Some studies adopted a cocurrent approach, in which both partners engaged separately with the same CA. For example, ParentText (Schafer et al., 2023, 2025) was designed to allow both parents to access the same agent individually, while LEAVES incorporated cocurrent features by enabling bereaved users to interact with content recorded by the deceased partner. PocketBot (Zheng et al., 2021) combined both cocurrent and conjoint modes by providing a shared couple chat room alongside private chat rooms for each partner. Finally, Li & Rau's (2021) IoT-CA was designed in a conjoint mode, enabling couples to interact with the system together.

### 5.1.3. Design Dimension 3. Implementation

*Emotional Competence*
This category focuses on the emotional detection and expression functions of CAs. Most studies did not directly implement features for recognizing or expressing emotions, though some incorporated limited emotion-related functions. For instance, Troitskaya and Batkhina (2022) asked users about their mood, Yuksel and Kocaballi (2022) provided responses to users' emotional expressions, and Zheng et al. (2021) encouraged emotional expression through their intervention. However, none of these studies implemented actual emotion recognition or expression capabilities. A few exceptions included Van Velsen et al. (2024), which monitored users' emotions to provide mood check and tailored reminders, and Oshio (2024), whose divorce negotiation chatbot analyzed users' emotions to deliver both legal and emotional support. The remaining studies (Li & Rau, 2021; Khatra et al., 2024; Mauricio et al., 2024; Schafer, Lachman, Zinser et al., 2025; Schafer, Lachman, Gardner et al., 2023; De Filippo et al., 2023; Njogu et al., 2023) did not report features related to emotion detection or expression.

*Responding to users' emotional expression*
Beyond emotion recognition and expression, the way CAs responded to users' emotional disclosures was also examined. Troitskaya and Batkhina (2022) did not directly address emotions but instead asked users to reflect on their feelings. Yuksel and Kocaballi (2022) employed validation strategies to acknowledge emotional expressions. Zheng et al. (2021) functioned primarily as a messenger, delivering emotional content to the partner rather than responding directly. Although they did not provide explicit responses, Van Velsen et al. (2024) and Oshio (2024) monitored or analyzed users' emotional states. The other studies (Li & Rau, 2021; Khatra et al., 2024; Schafer, Lachman, Zinser et al., 2025; Schafer, Lachman, Gardner et al., 2023; De Filippo et al., 2023; Njogu et al., 2023; Mauricio et al., 2024) did not incorporate responses to emotional expression.

*Interaction mode and technologies*
Most of the interventions relied on text-based interaction (Troitskaya & Batkhina, 2022; Yuksel & Kocaballi, 2022; Zheng et al., 2021; De Filippo et al., 2023; Schafer, Lachman, Zinser et al., 2025; Schafer, Lachman, Gardner et al., 2023; Njogu et al., 2023; Mauricio et al., 2024; Oshio, 2024). Van Velsen et al. (2024) used a web-based platform that combined text with visual elements, while Khatra et al. (2024) supplemented text with visual feedback. Li & Rau (2021) uniquely implemented voice-based interaction using a smart speaker.

Most text-based systems were deployed on existing messaging platforms such as Facebook Messenger (Zheng et al., 2021; Schafer, Lachman, Gardner et al., 2023), Telegram (De Filippo et al., 2023), and WhatsApp (Njogu et al., 2023), while some adopted multi-platform delivery, including combinations of WhatsApp, Telegram, Facebook Messenger, and SMS (Schafer, Lachman, Zinser et al., 2025; Mauricio et al., 2024). Several studies used proprietary platforms (Troitskaya & Batkhina, 2022; Yuksel & Kocaballi, 2022), and one study leveraged ChatGPT-4 (Oshio, 2024). Text-and-visual systems were delivered through proprietary or web-based platforms (Van Velsen et al., 2024; Khatra et al., 2024). The only voice-based deployment was implemented via a smart speaker (Li & Rau, 2021).

In terms of dialogue implementation technologies, many studies adopted a rule-based approach (Zheng et al., 2021; De Filippo et al., 2023; Schafer, Lachman, Zinser et al., 2025; Schafer, Lachman, Gardner et al., 2023; Njogu et al., 2023; Van Velsen et al., 2024; Mauricio et al., 2024; Khatra et al., 2024). Responsive algorithms were the next most common (Troitskaya & Batkhina, 2022). Li & Rau (2021) and Yuksel & Kocaballi (2022) employed simulation or Wizard-of-Oz methods. Notably, Oshio (2024) was the only study to implement generative AI.

### 5.2. Synthesis and Design Considerations

To identify the key considerations for designing CAs for couples, a thematic analysis was conducted. First, the intersection points among the characteristics of each dimension, organized by RQ, were examined. To support the emerging themes, the findings and UX insights from the twelve selected studies were extracted and reviewed. The extracted statements and results were then compared and analyzed alongside the study characteristics to identify the final themes. Three overarching themes were identified: First, the need for persona design grounded in a relational perspective. Second, State-of-the-art technologies for couples, representing more advanced and context-aware systems. Third, the importance of shifting from content-centered to user-centered service design.

### 5.2.1. The Need for Designing Personas that Reflect a Relational Perspective

This theme emerged from the recognition that most existing studies on CAs for couples focus on individual-oriented interventions and general interaction principles, rather than addressing the relational context unique to couples. It was also informed by unclear persona settings, limited responses to users' emotional expressions, insufficient understanding of relational theories, and interaction strategies that overlook the couple's dynamics. These gaps suggest that current CA designs fail to consider the emotional and reciprocal nature of couple relationships.

These interpretations were supported by user experience statements. Participants reported discomfort with interactions that lacked emotional empathy or human understanding (Yüksel & Kocaballi, 2022) and expressed the view that conversational agents could not replace human counselors (Van Velsen et al., 2024). Other statements requested interventions that would help resolve couple conflicts (Zheng et al., 2021) and emphasized the need for a professional, relationship-focused counseling persona (Yüksel & Kocaballi, 2022; Oshio, 2024). These findings suggest that persona settings should enable emotional bonding through dialogue and reflect the relational understanding characteristic of professional couple counselors.

In the field of couple and family therapy, the relational perspective has long been regarded as the foundation of therapeutic effectiveness. Sprenkle et al. (2009) analyzed the distinctive features contributing to successful couple and family therapy. They identified four key factors: (1) conceptualizing difficulties in relational terms, (2) disrupting dysfunctional relational patterns, (3) expanding the direct treatment system, and (4) expanding the therapeutic alliance.

Although there is not yet full consensus on the common factors of couple and family therapy, the professional competencies of relational counselors provide complementary insight. The American Association for Marriage and Family Therapy (AAMFT, 2004) developed a set of core competencies that continue to serve as a framework for training programs in the field (Gehart, 2011). These competencies are organized into six primary domains and five secondary domains, emphasizing the assessment and diagnosis of problems from a relational-systemic perspective, the use of relationship-centered interventions, and the identification and modification of dysfunctional relational patterns.

Reflecting both perspectives, key elements relevant to CA design for couples include incorporating relational questioning, recognizing the dyadic therapeutic system, establishing a therapeutic alliance, and applying techniques that promote relational change in dysfunctional dynamics. These principles represent essential considerations for persona design in conversational agents intended for couple interventions.

The uniqueness of couples lies in their relational nature. Unlike interventions that focus on individual intrapsychic processes, addressing the relational dimension enhances therapeutic effectiveness. However, general interaction competencies should also be considered in persona settings. HCI literature highlights trust, rapport, active listening, and contextual understanding as essential elements of good conversation (Clark et al., 2019). Similarly, studies on healthcare and counseling CAs have noted that human-like communication and socially grounded personas can foster emotional bonding and self-disclosure (Nißen et al., 2022; Van Pinxteren et al., 2020). Therefore, persona design in CAs for couples is particularly important. A persona that integrates interactional competence with a relational perspective can better support couple-oriented interventions.

Accordingly, the following persona design considerations are proposed. Each persona was structured to reduce its therapeutic tone and focus instead on intervention-oriented functions.

*Role Definition: Relational Expert in the Couple Context*
The CA is designed to serve as a relational expert applicable to various couple-related issues. A relational expert focuses on reframing problems within a relational perspective and facilitates dialogue between partners. Couple-related topics are diverse (Lebow & Snyder, 2022). Accordingly, the CA may intervene in different ways: promoting mutual understanding in relationship enhancement, helping victims of high-risk IPV situations recognize dysfunctional relational patterns and establish personal boundary strategies, and supporting individuals after relationship dissolution in processing unresolved emotions and understanding the meaning of loss. Reframing from a relational perspective thus involves not only understanding the partner but also encouraging self-reflection within the relationship system. In conversational strategy, the CA adopts circular rather than linear questioning to promote relational thinking and broaden users' understanding of their interactions.

*Personality: A Warm and Trustworthy Expert*
The persona's personality is recommended to be designed around core therapist qualities such as empathy, non-judgment, and authenticity. These attributes are fundamental to establishing a therapeutic relationship in couple and family therapy (Petterson, 1987; Sprenkle, 2004). However, while such traits are important, it is equally essential that the persona demonstrate a clear understanding of relational dynamics. The CA should also maintain explicit ethical boundaries (AAMFT, 2004). It may provide emotional support within the relational context but should not engage in therapeutic practice. In cases involving crisis situations or users expressing psychological distress, the CA must be able to refer users to qualified human professionals. These boundaries should be communicated transparently to users. In addition to personality, providing contextual attributes such as name, age, or professional background can enhance the CA's perceived trustworthiness and professionalism (Nißen et al., 2022).

*Position Within the Relationship: Role According to Treatment Mode*
Services targeting couples may be directed toward individuals or dyads. In such contexts, the CA should maintain an appropriate stance by understanding the relational flow. Reflecting the common factors of couple and family therapy, the CA's role incorporates *expanding the direct treatment system* and *expanding the therapeutic alliance* (Sprenkle et al., 2009), thereby sustaining awareness of the couple relationship and forming a 2:1 alliance. The design also considers management of multiple alliances, as well as sensitivity to cultural, gender, and power differences within relationships.

In couple therapy, treatment modes are typically categorized as *individual*, *conjoint*, and *concurrent* (Sperry, 2018). For designing the CA's position within the relationship, the conjoint mode requires particular attention. When both partners interact simultaneously with the CA, it should function as a neutral mediator. To achieve this, the CA can be designed to operate based on *multiple alliances* (Sprenkle, 2004). For example, when communicating with both partners, the CA should avoid aligning with one partner by asking reflective questions that include the other's perspective. Furthermore, it should demonstrate cultural sensitivity by adjusting its dialogue according to factors such as cultural background, gender dynamics, or regional context of the couple (Schafer, Lachman, Gardner et al., 2023).

### 5.2.2. State-of-the Art Technologies for Couples: Toward More Advanced Approaches
Unlike individual users, couples represent a unique relational unit composed of two interdependent agents. As a dyadic system, each partner's utterance has the potential to influence the other. Moreover, couples share an intimate and emotionally charged relationship, which requires that conversational agents integrate both aspects of emotional competence—detection and expression. Additionally, depending on the intervention context, the mode of engagement may vary, encompassing individual, concurrent, or conjoint approaches (Sperry, 2018). Therefore, intervention design should adhere to a relational rather than an individual-centered perspective. This relational specificity and distinctiveness suggest that such interventions could be effectively supported by state-of-the-art (SOTA) technologies.

Across each dimension, notable gaps were identified in the application of state-of-the-art (SOTA) technologies. Most existing studies rely on dialogue designs lacking contextual awareness, absence of mode-switching capabilities, limited emotional competence, and predominantly rule-based interaction structures. Building on these gaps, this study introduces emerging technologies from recent publications and preprints (2024–2025) that have the potential to support relational specificity. Based on these advancements, design considerations are proposed for each treatment mode.

Recent advances in conversational artificial intelligence have demonstrated significant progress through the development of large language models (LLMs). LLMs are trained on vast corpora of linguistic data, enabling them to understand context and generate coherent text. These models have facilitated a shift from single-turn to multi-turn dialogue systems, allowing for sustained and contextually informed interactions that retain and reference prior exchanges (Yi et al., 2024). Specifically, multi-turn dialogue systems can recall previous utterances to produce consistent responses, thereby maintaining both the emotional continuity and thematic coherence of a conversation. Such capabilities are particularly valuable for tracking emotional dynamics and preserving topic consistency within couple interactions.

The multi-turn dialogue structure is generally divided into task-oriented dialogue (TOD) and open-domain dialogue (ODD). Among these, ODD aims to facilitate natural and emotionally engaging interactions that do not seek predetermined answers but rather support empathy, immersion, and spontaneous emotional exchange (Yi et al., 2024). Consequently, ODD is considered a more suitable framework for couple-oriented conversational agents. With the integration of LLMs, ODD systems can infer user emotions from tone and word choice, generate empathic responses, and enable personalized dialogue experiences.

ODD systems are typically categorized into two types: retrieval-based and generative. Retrieval-based systems provide factual and evidence-based responses but lack flexibility, while generative systems offer more natural and adaptive dialogue yet risk producing inconsistent or inaccurate information. These limitations, however, are increasingly mitigated through the emergence of hybrid systems that combine the strengths of both approaches. Such advancements indicate a transition toward conversational technologies capable of overcoming previous interactional constraints.

Earlier conversational systems were predominantly text-based, but recent developments have introduced multimodal large language models (MLLMs) such as GPT-4V, Gemini 1.5, Claude 3, and LLaVA, which can process visual, auditory, and textual information simultaneously (Caffagni et al., 2024). These models integrate perceptual processing and linguistic reasoning within a unified conversational framework, enabling interactions that approximate human communication across multiple sensory modalities. In particular, voice-based real-time interaction has advanced considerably. For example, full-duplex spoken language models (FDSLMs) are capable of processing synchronous speech (Chen & Yu, 2025), allowing the system to perform natural conversational behaviors such as backchanneling and mid-turn interjections while listening to the user, closely mirroring authentic human dialogue.

Speech separation and speaker recognition technologies, which are essential for managing interactions involving multiple speakers, have become increasingly sophisticated in handling long-term conversational contexts. Recent research employing dual-path recurrent neural networks (DPRNN) and hybrid CNN-RNN architectures has enabled efficient processing of extended dialogues by combining target speaker extraction (TSE) with speaker diarization, allowing real-time identification of who spoke, when, and what was said (Li et al., 2025). These advancements present promising opportunities for conjoint modes, where accurately distinguishing between two users' utterances can support precise timing of interventions.

For couple-oriented conversational agents, emotional competence remains a fundamental requirement (Allouch et al., 2021). Recent developments in emotion recognition and expression technologies conceptualize emotion not as a single signal but as a composite pattern encompassing language, speech, facial expression, and gaze, thereby enhancing the agent's capacity for emotional understanding (Geetha et al., 2024). Multimodal emotion recognition (MER) systems integrate these multiple signals, using cross-attention fusion mechanisms to interpret emotional context with greater accuracy. For instance, the Empatheia model integrates textual, vocal, and visual modalities to generate empathic responses characterized by high emotional coherence across avatar expressions, vocal tone, and linguistic phrasing (Zhang et al., 2025). Such multimodal approaches indicate that conversational agents have progressed from merely detecting emotional states to expressing emotions authentically. The integration of emotion recognition and expression technologies provides a foundation for CAs to not only understand couples' dialogues empathically but also reframe them relationally to support emotional recovery.

In couple interventions, a critical consideration is the design of CA–dyadic interaction, which aligns technically with the domain of multi-party conversational agents (MPCAs). Conversational agents designed for multiple users have evolved from simply tracking what was said to discerning who spoke, with what emotion and intention. This capability allows the system to decompose couple dialogue, detect each partner's emotions, engagement levels, and communicative intentions, and accordingly adjust the timing of interventions (Sapkota et al., 2025).

Technically, these capabilities are supported by models that summarize or organize multi-user conversations, separate and align overlapping speech segments, and analyze logical relations between utterances. Moreover, research on multi-user conversational interfaces (MUCI) has advanced toward integrating verbal and nonverbal cues such as speech, facial expressions, gestures, gaze, and interpersonal distance to facilitate role-aware interaction, mediate negotiation and competition among users, and maintain shared conversational context (Wagner et al., 2025).

These technological developments can be directly applied to the design of concurrent and conjoint modes for couple-oriented conversational agents. Accordingly, the following section presents key technological design considerations derived from these insights.

*Individual Mode or Common Key Considerations*
Couple-oriented conversational agents should be designed not merely as information providers but as relational experts capable of understanding the complexity of couple dynamics and supporting professional-level interventions. To achieve this, the agent must learn the structure of relational questioning through fine-tuning and generate contextually tailored dialogues aligned with various relationship themes. This requires an LLM-based open-domain dialogue (ODD) structure, which enables the CA to guide conversations that focus on relational understanding rather than simple information exchange, adapting dialogue flow to the individual user's context.

Moreover, by incorporating past conversation histories and emotional fluctuations, the intervention can evolve into a continuous process rather than a one-time exchange. Context-based dialogue structured around relational questions can foster relational insight for users while also allowing the system to detect potential crises such as emotional distress, conflict, or signs of abuse. This capacity contributes to enhancing the CA's role in ensuring couple safety.

The conversational form should be emotion-centered, reflecting the persona's empathic and non-judgmental nature. To accomplish this, the CA must recognize the user's current emotional state and determine how to respond in a manner that resonates effectively. Individual participation in couple interventions often involves emotionally charged concerns related to the relationship, which require sensitive emotional handling. Through multimodal fusion, the CA can integratively perceive the emotional context and express responses through multiple sensory channels, thereby supporting richer and more authentic relational engagement.

*Concurrent Mode Key Considerations*
The central consideration in the concurrent mode lies in managing information sharing between partners. In this mode, each partner interacts separately with the same conversational agent, making it essential to guarantee user autonomy over what information is shared. When users discuss their perspectives individually with the CA, selectively sharing portions of these experiences can facilitate mutual understanding and promote relational insight.

To support this process, dialogue summarization models such as SummN, Hierarchical Attention, and TOFUeval can be employed to generate concise and intuitive summaries of individual conversations. These summaries can then be reformulated into formats that naturally bridge the two partners' dialogues. Furthermore, integrating the semantic inference capabilities of LLMs with the multimodal perception of MLLMs enables the generation of relational summaries that go beyond textual compression to include emotional context, thereby providing more effective relational interventions.

For example, by analyzing nonverbal cues such as vocal tone, facial expression, gaze, and linguistic patterns, the CA can interpret the emotional implications of each utterance and paraphrase the summarized content into empathetic language that each partner can receive constructively. Additionally, allowing users to review and confirm the LLM-generated summaries before sharing them strengthens user autonomy and trust in the system.

*Conjoint Mode Key Considerations*
In the conjoint mode, the conversational agent must incorporate functions that account for multiple alliances between the CA and the dyad. Specifically, the system should be capable of identifying the shared context within the conversation and mediating conflicts that arise between partners (Wagner et al., 2025). Shared context can be analyzed through speaker separation and diarization technologies, which enable the system to differentiate between speakers, analyze conversational flow, and map the structure of interactions. These features allow for the identification of speaking frequency, interruptions, and turn-taking balance, ensuring that neither partner becomes excluded from the dialogue and that both are given equitable opportunities to express themselves.

Building upon these techniques, multimodal integration of verbal and nonverbal data can be used to model commonalities and discrepancies in partners' expressions. Such modeling helps couples recognize points of agreement or misunderstanding that may otherwise go unnoticed during conversation. Additionally, conflict mediation can be enhanced through adaptive user profiling, where the CA generates personalized profiles based on prior interactions and adjusts its responses accordingly (Wagner et al., 2025).

When integrated with emotion recognition capabilities, these features become particularly meaningful. The system can be trained to detect when emotional regulation is compromised, prompting the CA to issue reminders or suggest a temporary pause in dialogue. This aligns with therapeutic principles recommending brief breaks of around twenty minutes when emotional arousal exceeds manageable levels (Gottman & Silver, 2012). Through such mechanisms, the CA can play an active role in maintaining constructive communication and emotional safety within conjoint interventions.

### 5.2.3. Not Content-Centered, But Relationship-Centered Design

This section draws inspiration from the findings of Schafer's studies (2023; 2025). While Schafer's research provided rich and meaningful content, it also reported low user engagement, short usage duration, and limited completion rates. The gap between high-quality content and sustained usability represents a critical challenge in designing CA services for couples.

Although the selected studies employed diverse research designs, those based on user evaluations primarily emphasized content usefulness and functional convenience. While these aspects are indeed important for assessing user experience, they are insufficient for contexts involving relational complexity such as couples. For such contexts, ethical integrity and the quality of systemic interaction should be regarded as more fundamental design considerations than content-centered usefulness.

Across dimensions, the key gaps identified include a lack of transparency and limited personalization and adaptability to couple-specific contexts. These shortcomings highlight the need for conversational agents that can ethically and dynamically respond to relational nuances within couple interventions.

User evaluation results across the selected studies were largely positive regarding content and functionality, with only a small portion of critical feedback or user requests. However, because most evaluation criteria focused on assessing content quality or technical performance, this study also examined whether ethical design and systemic interaction were adequately reflected across dimensions. In other words, beyond assessing what was provided, the analysis sought to identify gaps in how and under what relational and ethical conditions the interventions were delivered.

User feedback revealed recurring concerns such as cultural misalignment or mismatched content relative to relationship stage (Zheng et al., 2021), excessive or overly condensed information (Njogu et al., 2023), and privacy concerns even among users in non-clinical or low-risk relationships (Zheng et al., 2021; Yuksel & Kocaballi, 2022). These findings provide meaningful insights for designing couple-oriented CA systems. They suggest that although many studies have focused on improving technical performance and content completeness, the relational quality and ethical aspects of users' experiences have received comparatively less attention.

This pattern can be understood as a natural phase in the early development of CA technologies, when establishing a technical foundation was the primary goal. Nevertheless, in couple-specific contexts, one partner's experience directly affects the other's emotional well-being and relational stability, so functional efficiency alone is not sufficient. The objective of a couple-oriented CA should go beyond individual satisfaction to ensure the psychological safety and relational balance of both partners.

From this perspective, user-centered design should be understood not merely as a way to enhance user experience but as an ethical approach aimed at protecting user autonomy and relational safety. In this sense, function-centered design can be seen as a necessary condition for technological progress, whereas user-centered design represents a sufficient condition for achieving relational and ethical maturity.

To advance this view, the present discussion integrates ethical principles from both CA and UX design frameworks and from systemic ethics in couple therapy. CA and UX ethics emphasize five main principles: transparency, which requires that users understand the agent's identity, operation, and scope of involvement; autonomy and privacy, which ensure users' control over data collection and storage; trust; and accountability, which clarifies responsibility and oversight in the event of system errors (Wambsganss et al., 2021; Dubiel et al., 2022; Rahsepar Meadi et al., 2025).

In contrast, systemic ethics highlight client autonomy, confidentiality, safety, boundary management, and therapist positionality. The latter recognizes that therapists cannot remain entirely neutral and that their values inevitably influence the therapeutic process, representing a distinctive relational ethical stance (Gottlieb et al., 2008; Murphy & Hecker, 2020).

Although these two perspectives focus on different aspects of ethics, one centered on the system and the other on relationships, they share the same goal: to ensure a high-quality, ethically grounded experience for users or clients. This section integrates both perspectives to propose design considerations that reflect ethical responsibility in relationally oriented CA interventions for couples.

*Relational Ethics–Informed Design Proposals*
Building on the insights discussed above, this study proposes five design principles for conversational agents intended for couple users.

First, ethical transparency requires that the scope of data collection and intervention be explicitly communicated, including clear guidance on what data are stored or not stored within the service. This transparency should specify both the boundaries of information sharing within the couple and the extent to which data are disclosed outside the couple.

Second, boundary management involves preventing users from becoming overly immersed in or dependent on interactions with the CA. At the start of engagement, the CA should clearly define the limits of its involvement. For example, in cases involving high-risk situations such as abuse or severe psychological distress, users should be informed that professional therapeutic intervention is necessary alongside the CA's support.

Third, relational stage-sensitive intervention should be incorporated into design. This principle emphasizes that interventions must vary according to the relational stage, which includes not only the duration of the relationship but also its current risk level. For users experiencing high-risk circumstances, such as reports of abuse or mental harm, the CA should be designed to pause the intervention and connect users to appropriate professional services. For lower-risk couples, the system should assess the relationship's duration and context to provide tailored, stage-appropriate support.

Fourth, self-reflexive design draws from the systemic concept of therapist positionality. Designers, like therapists, are not neutral; their values inevitably influence content and interaction design. While CAs do not possess human emotions, beliefs, or values, the philosophical assumptions of their creators are embedded in their initial design and can become more pronounced as the CA's learning capacity evolves. Therefore, designers should engage in multidisciplinary consultation and make their guiding values and assumptions transparent.

Fifth, dyadic user experience calls for understanding the sociocultural backgrounds of both users and ensuring conversational balance and fairness between them. As discussed in previous sections, this involves designing for multiple alliances within the dyad while going beyond simple fairness to include sensitivity to cultural diversity and cross-cultural awareness.

## 5.3. Final Design Framework
Based on the themes derived, final design considerations were identified. These considerations are categorized into eight core components essential for designing conversational agents for couples: (1) agent persona, (2) individual mode, (3) concurrent mode, (4) conjoint mode, (5) ethics, (6) data and privacy, (7) interaction pattern, and (8) safety.

Each component was organized according to the technical and relational gaps identified in the preceding thematic analysis. They encompass the agent's role and characteristics, the application of technologies across different modes, and design directions concerning relational ethics and user safety.

Detailed design principles and technical application points are comprehensively illustrated in [Table 2]. This figure provides an integrated visualization of the key design goals for each component and their relational implications within the couple context.

*[ Table 2. Final Design Framework ]*

| Categories | Design Focus | Key Considerations |
|---|---|---|
| 1. Agent Persona | Acting as a relational expert | - Role: Relationally Experts<br>- Characteristic: empathic, non-judgmental, ethically bounded<br>- Maintaining neutrality within multiple alliances |
| 2. Individual Mode | Reframing individual dialogue through a relational lens | - LLM-based ODD multi-turn conversation<br>- Multimodal-based emotion detection and expression |
| 3. Concurrent Mode | Ensuring autonomy in shared information between partners | - Summarization-based relational sharing<br>- Emotionally contextualized summaries<br>- User approval prior to data sharing |
| 4. Conjoint Mode | Managing real-time multi-party interaction | - Speaker separation and identification (Diarization)<br>- Turn-taking and overlap regulation<br>- Cool-down suggestions in emotionally heightened situations |
| 5. Ethics | Ensuring relational ethics and transparency | - Ethical transparency<br>- Boundary management<br>- Relational stage-sensitive intervention<br>- Self-reflexive design<br>- Dyadic user experience |
| 6. Data & Privacy | Protecting personal data and ensuring user control | - Minimal data collection<br>- Explicit consent<br>- Granular intra-couple sharing (sentence / topic / emotion level)<br>- User review & withdrawal rights<br>- Audit log visualization (data flow, model reasoning) |
| 7. Interaction Pattern | Establishing emotion-centered dialogue structures | - Empathic and non-judgmental language<br>- Relational question frameworks<br>- Integration of prior dialogue and emotional shifts<br>- Sustained engagement |
| 8. Safety Mechanism | Detecting and responding to crises | - Detection of risk signals (abuse, self-harm, or partner-harm)<br>- Referral to human experts<br>- Disclosure of limitations and potential errors |

## 6. Discussion

This systematic review analyzed studies on CA designed for couple relationship interventions and derived an integrated foundation for design that encompasses technical, relational, and ethical perspectives. The findings indicate that most existing CA for couples remain focused on individual-centered interventions and rule-based interactions, with limited reflection of relational context and emotional dynamics. These results highlight the necessity of approaching design from the perspective that couples are not merely two users but an interdependent relational system.

First, the analysis revealed that in most previous studies, the role and persona of the agent were only partially defined. While attributes such as name or personality were often included, relational expertise and ethical boundary-setting were rarely incorporated. This limitation suggests that designs which fail to capture the relational characteristics of couple's risk generating impersonal or non-empathic experiences. Future CAs for couples should therefore adopt the persona of a relational expert, incorporating dialogue structures that encourage relational questioning and circular thinking. This direction is grounded in the common factors approach proposed by Sprenkle et al. (2009) and the professional competency framework of AAMFT (2004), indicating the potential for technically modeling the core relational mechanisms used by human therapists to promote change.

Second, in terms of technological implementation, most studies relied on text-based and rule-based interactions, with limited capabilities for emotion recognition and expression. Functions that could recall prior sessions or integrate previous conversational content were generally absent. This indicates that current couple-oriented CAs lack contextual continuity, which is essential for addressing ongoing relational change. Future designs should integrate

LLM-based multi-session dialogue structures, speaker identification and turn-taking balance analysis, and multimodal emotion recognition technologies to sustain relational flow. Such developments point to the possibility of transforming CA from a simple responder into an interactive regulator that can dynamically mediate dialogue between partners.

Third, from a user experience perspective, the main issue identified was the limitation of content-centered design. User evaluations are predominantly focused on usefulness and convenience, with little attention to emotional safety or ethical transparency. Because couple relationships inherently involve reciprocal emotional influence, user satisfaction alone is not a sufficient measure of success. To address this, the present study integrated ethical principles from CA and UX design with systemic ethics from couple and family therapy to propose a relational ethics–informed design framework. This framework emphasizes technological transparency, boundary management, sensitivity to relational stage, designer self-reflexivity, and dyadic user fairness as core axes of ethical and relational design, thereby providing concrete criteria for ensuring user protection and trust in couple contexts.

Overall, these findings suggest that conversational agents in couple interventions should not function merely as supportive tools but as active relational support companions that mediate and facilitate relational change. By connecting relational theories in couple therapy with design principles from HCI, this study offers a new theoretical framework for understanding how technology engages with relationships. It contributes theoretically by extending the common factors theory of couple therapy into a technological context and practically by presenting eight core design components that can guide developers, clinicians, and researchers in collaborative design efforts.

Several limitations should be acknowledged. First, the number of studies included was limited, with most involving short-term interventions or small-scale experiments, making it difficult to assess long-term relational outcomes. Second, because the review focused on English-language studies, cultural diversity and non-Western relational patterns were underrepresented. Third, discussions of emotion recognition and multi-user interaction technologies remained theoretical, lacking empirical implementation and usability validation.

To address these limitations, future research should expand in three directions. First, relational questioning techniques from couple therapy experts should be systematized to build a relational question dataset that can be used for fine-tuning LLMs, thereby enhancing the agent's relational language competence. Second, experimental studies involving multi-user interactions are needed to empirically examine conversational balance, emotion regulation, and information sharing between couples and CAs in real-time dialogue settings. Third, the development of relational ethical fitness metrics would allow for the evaluation of ethical transparency and emotional safety, extending assessment beyond functional performance measures.

In conclusion, this study provides a systematic synthesis of the technical, theoretical, and ethical conditions required for conversational agents to address couple relationship challenges and proposes an integrated design framework. This holistic approach offers a foundational academic basis for the evolution of AI interventions from individual-centered therapy toward relationship-centered design.